\def\year{2026}\relax

\documentclass[letterpaper]{article} 
\usepackage{aaai25}  
\usepackage{times}  
\usepackage{helvet}  
\usepackage{courier}  
\usepackage[hyphens]{url}  
\usepackage{graphicx} 
\usepackage{natbib}  
\usepackage{caption} 
\usepackage{xspace}
\usepackage{makecell}
\usepackage{footmisc}
\DeclareCaptionStyle{ruled}{labelfont=normalfont,labelsep=colon,strut=off} 
\usepackage{algorithm}
\usepackage{algorithmic}
\usepackage{booktabs}
\usepackage{subfigure}
\usepackage{tcolorbox}

\tcbuselibrary{skins,breakable}
\newcommand{\xhdr}[1]{\vspace{1mm}\noindent\textbf{#1}}

\usepackage{xcolor}

\newcommand{\websiteA}{Website A\xspace}
\newcommand{\websiteB}{Website B\xspace}
\newcommand{\fourchan}{4chan\xspace}
\newcommand{\ncii}{NCII\xspace}
\newcommand{\ibsa}{IBSA\xspace}

\newcommand{\sncei}{SNCEI\xspace}
\newcommand{\archivedmoe}{Archived.Moe\xspace}

\newtcolorbox{findingbox}{
  enhanced,
  breakable,
  colback=gray!10,
  colframe=gray!40,
  boxrule=0pt,
  arc=4mm,
  left=4mm,
  right=4mm,
  top=2mm,
  bottom=2mm
}

\usepackage{newfloat}
\usepackage{listings}
\usepackage{threeparttable}
\floatstyle{ruled}
\newfloat{listing}{tb}{lst}{}
\floatname{listing}{Listing}
\title{Deepfake Pornography is Resilient to Regulatory and Platform Shocks}

\author{
  Alejandro Cuevas and
  Manoel Horta Ribeiro
}

\affiliations{
  Princeton University\\
  cuevas@princeton.edu, manoel@princeton.edu
}

\begin{document}

\maketitle

\begin{abstract}
Generative artificial intelligence tools have made it easier to create realistic, synthetic non-consensual explicit imagery (popularly known as deepfake pornography; hereinafter \sncei) of people. Once created, this \sncei is often shared on various websites, causing significant harm to victims. 
This emerging form of sexual abuse was recently criminalized in the US at the federal level by S.146, the \textit{TAKE IT DOWN} Act. A week after the bill's passage became effectively imminent, the MrDeepfakes website---one of the most notorious facilitators of \sncei creation and dissemination---shut down.
Here, we explore the impact of the bill's passage and the subsequent shutdown as a compound intervention on the dissemination of \sncei. 
We select three online forums where sexually explicit content is shared, each containing dedicated subforums to organize various types of sexually explicit content. By leveraging each forum's design, we compare activity in subforums dedicated to \sncei with that in other pornographic genres using a synthetic control, quasi-experimental approach.
Across websites, we observed an increase in the sharing and requests for \sncei, and, in some cases, in new contributors. These results indicate that the compound intervention did not suppress \sncei activity overall but instead coincided with its redistribution across platforms, with substantial heterogeneity in timing and magnitude. Together, our findings suggest that deplatforming and regulatory signals alone may shift where and when \sncei is produced and shared, rather than reducing its prevalence.
\end{abstract}

\section{Introduction}

\begin{figure}[t]
    \centering
    \includegraphics[width=\linewidth]{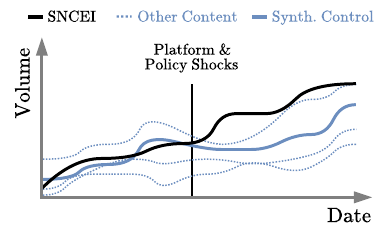}
    \caption{\textbf{Motivation.}
    We study the impact of the \textit{TAKE IT DOWN Act} and the deplatforming of MrDeepFakes on the production of synthetic non-consensual explicit imagery (\sncei) in three other websites known to host such content. Within these websites, we use the volume of non-\sncei-related content (dotted blue in the figure) to derive a counterfactual expected volume of \sncei had the shocks not occurred (the synthetic control; solid blue). We contrast it with the observed sharing of \sncei content (solid black), finding that \sncei sharing was resilient to this compound intervention, with significant spillovers to other platforms.}
    \label{fig:placeholder}
\end{figure}

Across historical contexts, individuals have produced defamatory or humiliating sexual portrayals (ranging from caricatures to manipulated photographs) intended to misrepresent identifiable targets without their consent~\cite{lake19thCenturyMan2021}. Over time, however, technological advances have lowered barriers to producing such content: from early darkroom `composite' techniques to mass-market image editing software like Adobe Photoshop. Most recently, advances and the popularization of generative artificial intelligence (GenAI) have increased the scale, realism, and accessibility of synthetic non-consensual explicit images (popularly known as deepfake pornography; hereinafter \sncei). 

The advancement and popularization of GenAI has transformed a niche or labor-intensive practice into a widely accessible one~\cite {gibsonAnalyzingAINudification2025,mehtaDeepfakes2023,hawkinsDeepfakesDemandRise2025} and enabled a growing ecosystem of online services dedicated to the creation of \sncei, like nudification websites~\cite{gibsonAnalyzingAINudification2025}, as well as platforms that facilitate their aggregation and distribution, e.g., MrDeepfakes~\cite{hanCharacterizingMrDeepFakesSexual2025}.
Under this new regime, fabricated sexual imagery has increasingly targeted non-public individuals, often within proximate social contexts such as schools and the workplace~\cite{weiWereUtterlyIllprepared2025,silvermanDressYourBuddys}. 
Victims of \sncei, who are disproportionately women~\cite{franksCriminalization2024}, are extorted and silenced and experience significant emotional distress, humiliation in their immediate social circles, harassment, and stalking as the intimate images are often accompanied by identifying information~\cite{bates2017revenge,franksCriminalization2024,weiWereUtterlyIllprepared2025}.

In response to the rapid expansion and documented harms of \sncei, policymakers have pursued regulatory interventions. In the US, Senate bill S.146 (the \textit{TAKE IT DOWN} Act) proposes federal prohibitions on the non-consensual online publication of intimate visual depictions, including authentic and computer-generated content, along with criminal penalties and takedown obligations for covered platforms~\cite{takeitdown_act_2025}. 
The passage of the bill on April 28, 2025, by the U.S. House of Representatives marked a salient regulatory signal, as President Trump had previously indicated he would sign the bill after reception~\cite{pressreleaseTakeItAct2025}, indicating imminent federal enforcement.

Within a week of the Act's passage by the US House of Representatives, MrDeepfakes---one of the largest known hubs for \sncei distribution~\cite{hanCharacterizingMrDeepFakesSexual2025}---announced its shutdown, citing that a ``critical service provider terminated service permanently~\cite{szetoThisCanadianPharmacist2025,wiseMajorDeepfakePorn2025}.%
\footnote{\citet{wiseMajorDeepfakePorn2025} suggests a connection with the bill's passage and the termination of services. 
} 
Together, these closely timed developments constitute a policy and platform-level intervention that plausibly altered both the production of \sncei and its migration across adjacent online ecosystems.

\xhdr{Present work.} 
In this paper, we causally examine how regulatory signals and the shutdown of a large platform dedicated to \sncei affect activity (i.e., new posts, new contributors, and new requests) across adjacent online websites. Exploiting the closely timed passage of the \textit{TAKE IT DOWN} Act by the US House of Representatives and the subsequent shutdown of MrDeepfakes as a compound intervention, we employ a synthetic control design (a commonly used quasi-experimental design) to estimate changes in content production and contributor activity on other related platforms.

We treat these closely timed developments as a salient \textit{compound shock} to the \sncei ecosystem and ask how production and participation reconfigure in response. Concretely, we use the House passage of the \textit{TAKE IT DOWN} Act and the subsequent deplatforming of MrDeepfakes as an event and examine whether adjacent websites exhibit (i) \textit{suppression}---a sustained decline in \sncei activity relative to a counterfactual baseline---or instead (ii) \textit{displacement and reconstitution}---a shift in activity to alternative services, potentially coupled with new participant entry.

To investigate this question, we assemble longitudinal traces from three websites that host \sncei, \fourchan, and Websites A and B (which we refer to using monickers for ethical and safety reasons).  We capture weekly activity related to (i) new posts in subforums dedicated to \sncei, (ii) new unique contributors, and (iii) new requests for generating \sncei. We then construct counterfactual trajectories for each outcome using synthetic control models fitted to the pre-intervention period, allowing us to estimate how each platform would have evolved in the absence of the combined regulatory signal and deplatforming event.

\xhdr{Results.} We find that (i) \sncei sharing activity does not exhibit a sustained decline following the compound intervention. Across all sites we study, post-intervention outcomes remain at or above their synthetic counterfactuals. Instead, adjacent platforms experience \textit{increases} in \sncei-related activity along multiple margins, including new posts and requests, and in some cases newly active contributors. (ii) We further observe that activity increases are not transient. Each website exhibits statistically significant and prolonged growth following the intervention window. Finally, (iii) responses differ across platforms: \websiteA shows increased posting and contributor activity prior to the interventions, whereas \fourchan and \websiteB experience increases only after the interventions. However, across all platforms, we found no evidence that take-downs were taking place.

\xhdr{Implications.} Our findings suggest that deplatforming a major \sncei hub---even alongside an imminent regulatory signal---primarily \textit{reallocates} activity across the ecosystem rather than suppressing it, showcasing its \textit{resilience}. Importantly, this reallocation is neither simultaneous nor uniform: some platforms experience increased activity prior to the intervention, while others see surges only afterward. This heterogeneity in timing and magnitude indicates that users may respond to regulatory change in stages, potentially reflecting differences in awareness, expectations about enforcement, or perceived platform risk.
For policies such as the \textit{TAKE IT DOWN} Act, these dynamics imply that, for many users and the websites we studied, awareness exists, but deterrence does not. Future work could examine how user expectations and platform characteristics shape these migration patterns and whether more active enforcement can better mitigate the displacement effect.

\section{Related Work}
\label{sec:related-work}

\xhdr{Image-Based Sexual Abuse.}
Past work on Image-Based Sexual Abuse (\ibsa), the umbrella term that encompasses \sncei, has focused on the motivations to produce \ncii content~\cite{hanCharacterizingMrDeepFakesSexual2025,timmermanStudyingOnlineDeepfake2023,gibsonAnalyzingAINudification2025,brighamViolationMyBody2024,kshetri2023},%
\footnote{There is a lack of definitional consensus on terminology. We discuss this in more depth in Appendix~\ref{apx:terminology}}
as well as the harms~\cite{weiWereUtterlyIllprepared2025,franksCriminalization2024,trifonovaMisinformationFraudStereotyping2024,mathews2024}, social perception~\cite{umbachNonConsensualSyntheticIntimate2024,brighamViolationMyBody2024,gamageAreDeepfakesConcerning2022}, dissemination~\cite{hanCharacterizingMrDeepFakesSexual2025,timmermanStudyingOnlineDeepfake2023}, and regulation~\cite{citronCriminalizing2014,franksCriminalization2024}. Most similar to our work, \citet{timmermanStudyingOnlineDeepfake2023} study the rise of MrDeepfakes following the deplatforming of \texttt{r/deepfakes}, and \citet{hanCharacterizingMrDeepFakesSexual2025} characterize the MrDeepfakes website after it became more established. In both cases, the authors find that women are disproportionately the victims; they, in turn, are subject to substantial mental and emotional harms~\cite{franksCriminalization2024}.

Financial gain seems to be one of the primary motivators for perpetrators~\cite{indicatorAINudifiersContinue}. But there are also other documented motivations, such as curiosity, power, revenge, and the desire to demean and silence victims~\cite{weiWereUtterlyIllprepared2025,citronCriminalizing2014,franksCriminalization2024}. Surveyed individuals across various societies seem to overwhelmingly condemn the creation and dissemination of \ncii. Notably, some individuals are more lenient with consumption~\cite{umbachNonConsensualSyntheticIntimate2024,brighamViolationMyBody2024}. On the other hand, some people in online communities seem eager to help in the creation of \sncei~\cite{gamageAreDeepfakesConcerning2022}.

\xhdr{Interventions Aimed at Problematic Content.}
Prior work has examined monetization restrictions---such as pressuring payment processors or advertising networks---as a lever to curb problematic online content, but finds limited deterrent effects. Platforms and creators routinely adapt by substituting alternative revenue streams, including fringe advertising networks~\citep{papadogiannakisWelcomeDarkSide2025}, cryptocurrency-based payments after mainstream processor withdrawal~\citep{civitaiCreditCardPayments}, off-platform monetization~\citep{huaAltMonetization2022}, and moderation-evasion tactics that allow continued operation on major platforms~\citep{dawoud2026fiverr}.

Another class of interventions targets access and distribution infrastructure, either by reducing exposure to problematic websites (e.g., search demotion or deterrence messaging~\cite{thakurThinkTwiceYou2025}) or by pressuring intermediaries such as hosting providers, registrars, and CDNs to withdraw service~\cite{Han_Kumar_Durumeric_2022}. These measures can produce acute disruption: MrDeepfakes, for example, shut down on 5 May after a ``critical service provider terminated service permanently''\cite{szetoThisCanadianPharmacist2025,wiseMajorDeepfakePorn2025}. However, prior work suggests that such takedowns rarely yield ecosystem-level suppression. Communities often relocate, fragment, or shift toward adjacent platforms with weaker moderation\cite{horta2021platform,russo2023spillover,monti2023online}. Evidence from deplatforming more broadly is mixed: some bans reduce activity and toxicity on the originating platform~\cite{chandrasekharan2017you,jhaver2021evaluating}, while others document migration and sustained engagement elsewhere, often in less visible spaces~\cite{ali2021understanding,urman2022they}. This literature motivates examining whether disruption of a central node produces durable reductions or primarily reallocates harmful activity across the ecosystem.

\section{Background}
\label{sec:background}

\subsection{Legal Landscape}
In the United States, early regulation of \ibsa developed through a combination of state-level criminal statutes and limited federal provisions.  Beginning in the early 2010s, states slowly began enacting ``revenge porn'' or intimate image abuse laws, faced with an emergence of websites dedicated to the sharing of \ncii~\cite{citronCriminalizing2014}. 
As of 2023, 48 states had passed laws addressing \ibsa in some form~\cite{franksCriminalization2024}. However, the U.S. Congress repeatedly failed to enact uniform federal legislation, such as the SHIELD Act in September 2023 and the DEFIANCE Act in 2024~\cite{s3696_defiance_act_2024,s412_shield_act_2023}.

The emergence of GenAI tools heightened the urgency of national legislation. Recently, the executive branch has emphasized \ibsa as a priority harm in the broader governance of AI~\cite{eo_ai_2023}. These efforts ultimately led to the introduction of the \textit{TAKE IT DOWN} Act in the US Senate on February 13th, 2025~\cite{takeitdown_act_2025}: a federal prohibition on the  distribution of \ncii depictions---including AI-generated content---and requires platforms to promptly remove such material upon notice. The bill's passage by the US Congress made its enactment on May 19th all but certain. In anticipation of the enactment, AI companies and service providers began to adjust their policies. 

Parallel to these developments, the UK entered phased enforcement of the 2023 Online Safety Act (OSA), which imposes platform-level duties to assess and mitigate users' exposure to content that is illegal under existing criminal law. On March 3rd, 2025, Ofcom (the UK's Office of Communications; a communications regulator) announced its enforcement programme~\cite{ofcom_illegal_content_risk_assessment_enforcement_2025}, requiring a wide range of websites to submit an illegal content risk assessment by March 16th~\cite{ofcom_quick_guide_online_safety_risk_assessments_2025}.
All three websites we consider in this study would still need to comply with Ofcom's guidelines regarding the range of content they host. As of now, both \fourchan and \websiteB have received formal information notice by Ofcom~\cite{ofcom_4chan_investigation_2025,ofcom_age_assurance_enforcement_programme_2025}.

A key difference between the UK's OSA and the US's \textit{TAKE IT DOWN} Act is that the former centers on regulatory enforcement of platform duties tied to existing criminal law, whereas the latter establishes, within a single statute, both a unified federal prohibition on the distribution of \ncii---including \sncei---and a mandatory notice-and-takedown regime for covered platforms.

\subsection{Websites Under Study}

We study three prominent websites that host \sncei. Two are dedicated primarily to sexually explicit material and include one or more subforums focused on \sncei. As shown in Table~\ref{tab:website-sizes}, the cumulative number of \sncei posts and contributors on these sites exceeds that reported for MrDeepfakes~\cite{hanCharacterizingMrDeepFakesSexual2025}. The third site, \fourchan, is a large forum organized into topic-specific ``boards,'' only a subset of which host sexually explicit content.
For ethical reasons (Section~\ref{sec:ethics}), we do not publicly name two of the websites, following prior work on \sncei~\cite{gibsonAnalyzingAINudification2025} and on forums that facilitate intimate-partner surveillance~\cite{belliniSocalledPrivacyBreeds2021}. We do name \fourchan, as it is widely known and has been extensively studied in prior research~\cite{bernstein4chanAnalysisAnonymity2011,Hine_Onaolapo_2017}.

\xhdr{\websiteA} is a large forum dedicated to the sharing of sexually explicit media. \websiteA is organized in subforums, threads, and posts. A subforum corresponds to a high-level category (e.g., ``Fakes/AI/Deepfakes'') and threads predominantly concern a specific person (e.g., ``Taylor Swift''). In each thread, users contribute posts that may include media, such as images, videos, or links to file-sharing sites. More than 90\% of the posts we consider contain media, as posts without media contributions are only allowed in certain subforums. \websiteA displays aggregate statistics for the entire website; at the time of writing, it has 2,717,116 registered users, 251,423 threads, and 5,592,048 posts — corresponding to 48,522,876 images, videos, and external URLs that point to file-sharing services. The list of subforums we consider ($n=32$) is shown in Table~\ref{tab:subforums}.
\websiteA maintains a subforum dedicated to ``Fakes/AI/Deepfakes''; lifetime counts are shown in Table~\ref{tab:website-sizes} and counts for this study's interval in Table~\ref{tab:subforums}. Other popular subforums are dedicated to sharing content from paid platforms (e.g., OnlyFans) or aggregating content from adult content creators. The ``Fakes/AI/Deepfakes'' subforum restricts postings to depictions of individuals who are at least 18 years old and who meet one of several notoriety criteria. Given these criteria, the vast majority of content in this category can reasonably be characterized as \sncei.

\xhdr{\websiteB} is a large imageboard where users share sexually explicit content that has been in operation since at least 2006. \websiteB allows registered users to upload posts and tag them with specified categories. Each category acts a subforum where all displayed posts are those that match the high-level category. All posts contain either a few images/GIFs or thousands. The list of subforums ($n$=68) we included is shown in Table~\ref{tab:subforums}.
\websiteB organizes content similar to traditional pornographic websites (e.g., action, ethnicity, fetishes, etc.), as observed in Table~\ref{tab:subforums}. Recently, they introduced the ``AI-generated'' tag for content generated using AI tools ($n$=43,981 posts). There is also a ``Fakes'' tag ($n$=10,566 posts) for content that involves digitally altering an image to make it appear to be someone else. Lastly, the ``Celebrities'' tag is used for content depicting celebrities ($n$=23,688 posts). \websiteB also has a ``computer-generated'' tag which is primarily oriented towards 3D renders, rather than the more specific ``AI-generated'' category.

\xhdr{\fourchan} is an anonymous imageboard with minimal moderation, organized into topic-specific boards (e.g., Anime \& Manga, Technology). Each board contains threads, which in turn consist of individual posts that typically include text and may include images, GIFs, or short video clips. \fourchan has been widely studied in prior work, particularly its politics board (\texttt{/pol/})~\cite{jokubauskaiteGenerallyCuriousThematically2020} and its ``random'' board (\texttt{/b/})~\cite{bernstein4chanAnalysisAnonymity2011}. 
Among \fourchan's 13 adult content boards (Table~\ref{tab:subforums}), we focus on \texttt{/r/} (``Adult Requests''), a request-oriented board where users seek help locating, identifying, or generating sexually explicit material---including \ncii (e.g., leaks). In recent years, requests involving AI tools have become increasingly common (Section~\ref{sec:results}), making \texttt{/r/} a useful proxy for demand for \sncei. Over our study period, we collected 496{,}531 posts across 42{,}234 threads from \texttt{/r/}. 
Because threads and posts are ephemeral, we rely on the \archivedmoe archiver to retrieve historical content from NSFW boards. Due to \fourchan's anonymity, contributor identities are unavailable, and we therefore analyze activity at the post and thread level.

\section{Methodology}
\label{sec:methods}

\begin{table}[t]
\centering
\renewcommand{\arraystretch}{1.25}
\resizebox{\linewidth}{!}{%
\begin{tabular}{lccc}
\hline
 & \begin{tabular}[c]{@{}c@{}}\textbf{\websiteA}\\(Fakes/AI/Deepf.)\end{tabular}
 & \begin{tabular}[c]{@{}c@{}}\textbf{\websiteB}\\(AI, Celeb., Fakes)\end{tabular}
 & \begin{tabular}[c]{@{}c@{}}\textbf{MrDeepfakes}\\(All; \citet{hanCharacterizingMrDeepFakesSexual2025}\end{tabular} \\
\hline
Contribs. & 12{,}151 & 10{,}160 & 1{,}880 \\
Posts    & 104{,}418 & 359{,}400 & 48{,}734 \\
\hline
\end{tabular}}
\caption{\label{tab:website-sizes} Size comparison across platforms. Contributors refer to unique users who have posted at least one post containing media. For \fourchan, the actual number of unique contributors is not known, but it is consistently in the hundreds of thousands, based on their homepage's statistics.}
\end{table}

\subsection{Data Sources}
\label{sec:data-sources}

We use \websiteA, \websiteB, and \fourchan as data sources to measure activity related to \sncei. For \websiteA and \websiteB, we design a scraper to collect timestamped posts and their corresponding authors. For \fourchan, we leverage a third-party archiver to collect historical threads and posts: \archivedmoe. \fourchan's threads are ephemeral and often deleted after short periods, which is why third-party websites provide historical archives. Furthermore, \fourchan is anonymous so we cannot infer the number of new contributors. For the purposes of this study, we restricted our measurement period to November 1st, 2024, to September 30th, 2025 (48 weeks). For each website, we collect all available threads and posts for each subforum in our measurement interval. The list of subforums we considered for each website, along with the number of posts and unique contributors, is shown in Table~\ref{tab:subforums}. For each subforum, we create a weekly timeseries for new posts and newly active contributors, a metric we describe in Section~\ref{sec:synth-setup}.

\subsection{Ethical Considerations and Open Science}
\label{sec:ethics}

\subsubsection{Data Collection.}
\label{sec:data-collection}
In collecting the data for this study, we did not violate any websites' Terms of Service (ToS) or their \texttt{robots.txt} files. All data from \websiteB and \archivedmoe (the \fourchan archiver) are publicly accessible; most data from \websiteA are also publicly accessible. \websiteA hides some of its subforums (such as the ``AI/Fakes/Deepfakes'' subforum, which is accessible only to free users). Since their ToS does not prohibit scraping, we registered an account to collect data from these hidden subforums.
Our analyses focus on aggregate data (i.e., counts of posts and contributors). We do not collect any images or videos. Posts contained associated usernames, and we used these to count contributor activity. We do not attempt to de-anonymize individuals (neither subjects nor contributors). Usernames, however, may contain PII if users choose to share them. Most usernames are publicly visible, except for those in the hidden section of \websiteA. We acknowledge that scraping these websites increased server load and bandwidth usage. We minimize the number of requests and employ conservative limits (when not explicitly specified in the sites' \texttt{robots.txt} files), well below the expected organic traffic these websites likely handle given their user bases.

\subsubsection{Website Name Disclosure and Data Sharing.}
\label{sec:website-disclosure}
We seek to balance Open Science principles with the minimization of potential harms. Past work on \sncei communities and websites repeatedly grapples with the question of whether to publicly name the websites under study~\cite{gibsonAnalyzingAINudification2025,hanCharacterizingMrDeepFakesSexual2025,belliniSocalledPrivacyBreeds2021}. As discussed by \citet{gibsonAnalyzingAINudification2025}, naming the websites and sharing data can cause harm within and beyond the research community. These websites contain sexually explicit images and videos of people who may not have consented to having those images shared. Publicly naming these websites can increase traffic, improve their search rankings, and help new users find and engage with them. When assessing these risks, \citet{hanCharacterizingMrDeepFakesSexual2025} chose to name the MrDeepfakes website because ``it has been named directly in prior work and is the top search result for celebrity sexual deepfakes.'' In our case, the two websites we censor do not meet these criteria. At the same time, we recognize the need for reproducibility. Following the framework of \citet{gibsonAnalyzingAINudification2025}, we are releasing our data and code on Zenodo, with de-identified website names. Researchers may request the identities for \websiteA and \websiteB. We will assess requests by verifying the identity of the requesting researcher, confirming that the requested data is appropriate for the stated research purpose, ensuring that requesters subject to age-verification laws for explicit content are at least 18 years old, and consulting institutional or external advisors when a request cannot be clearly resolved internally. We ask readers to refrain from de-identifying these websites based on the metadata shared in this paper. The dataset and code can be found here: \url{https://doi.org/10.5281/zenodo.18460085}.

\begin{figure*}[t]
\centering
\includegraphics[width=0.32\textwidth]{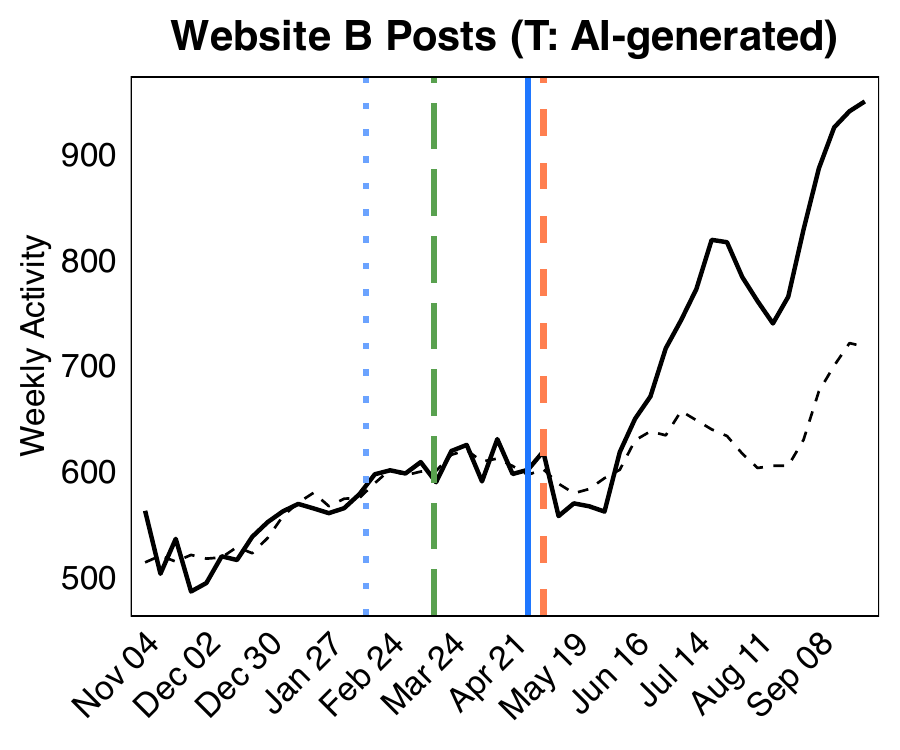}%
\includegraphics[width=0.32\textwidth]{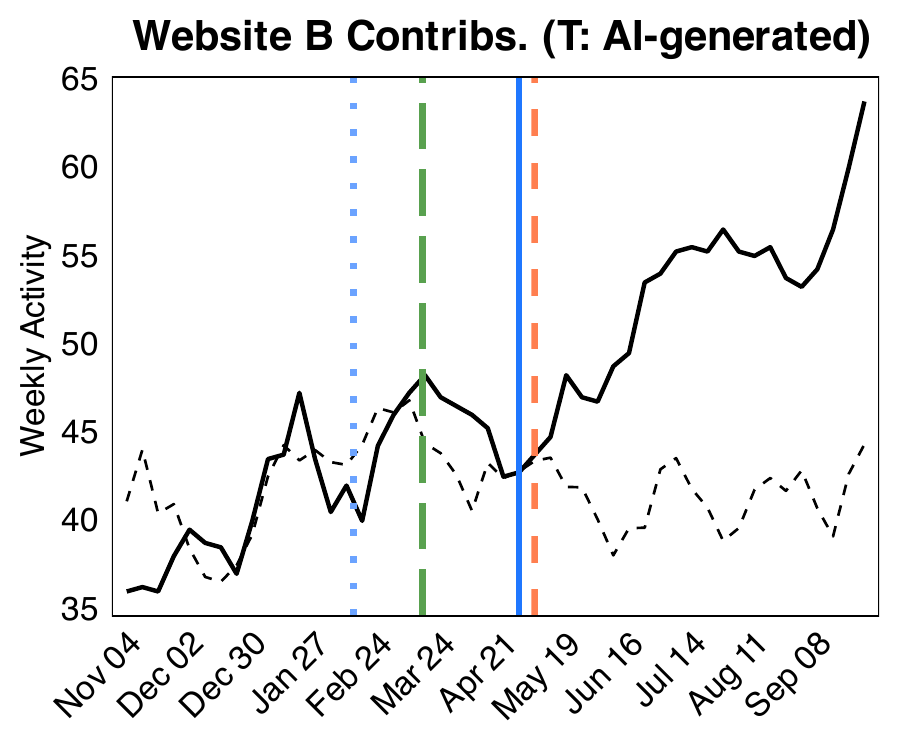}%
\includegraphics[width=0.32\textwidth]{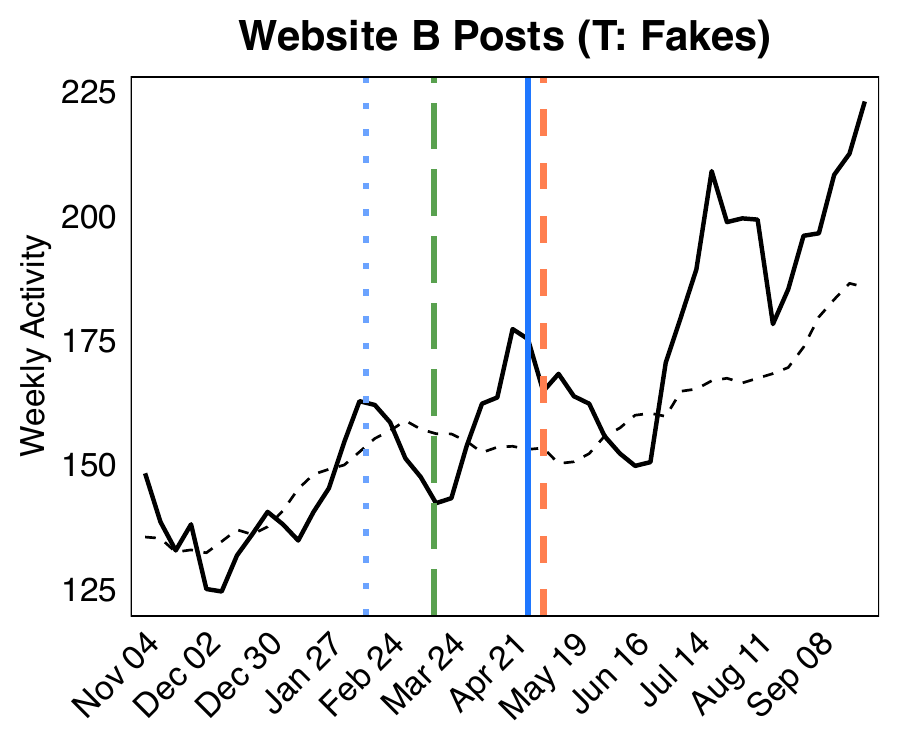}\\[2mm]
\includegraphics[width=0.32\textwidth]{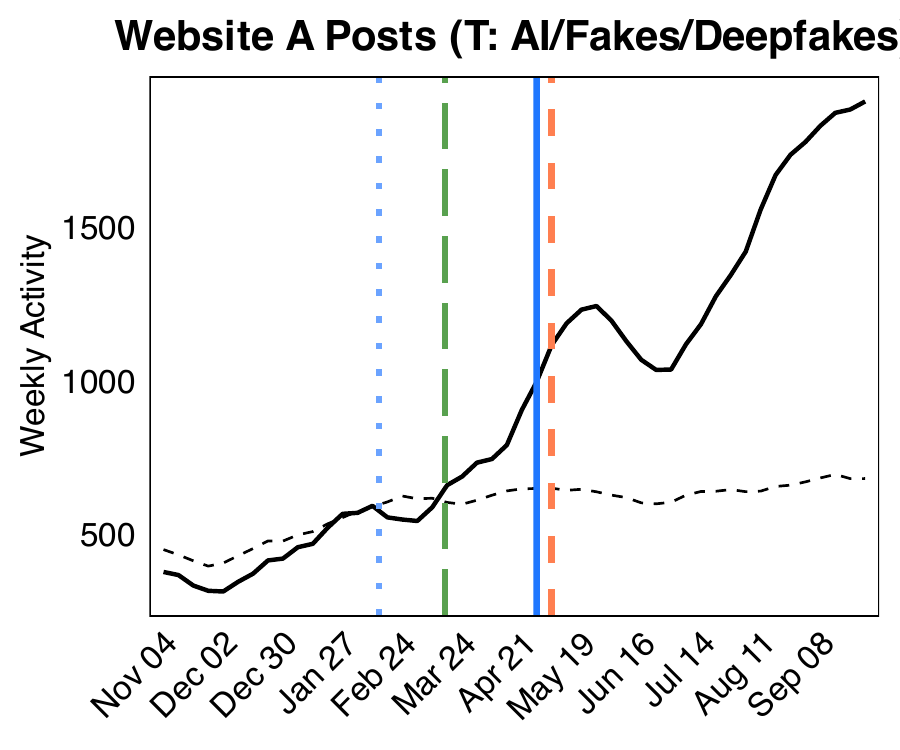}
\includegraphics[width=0.32\textwidth]{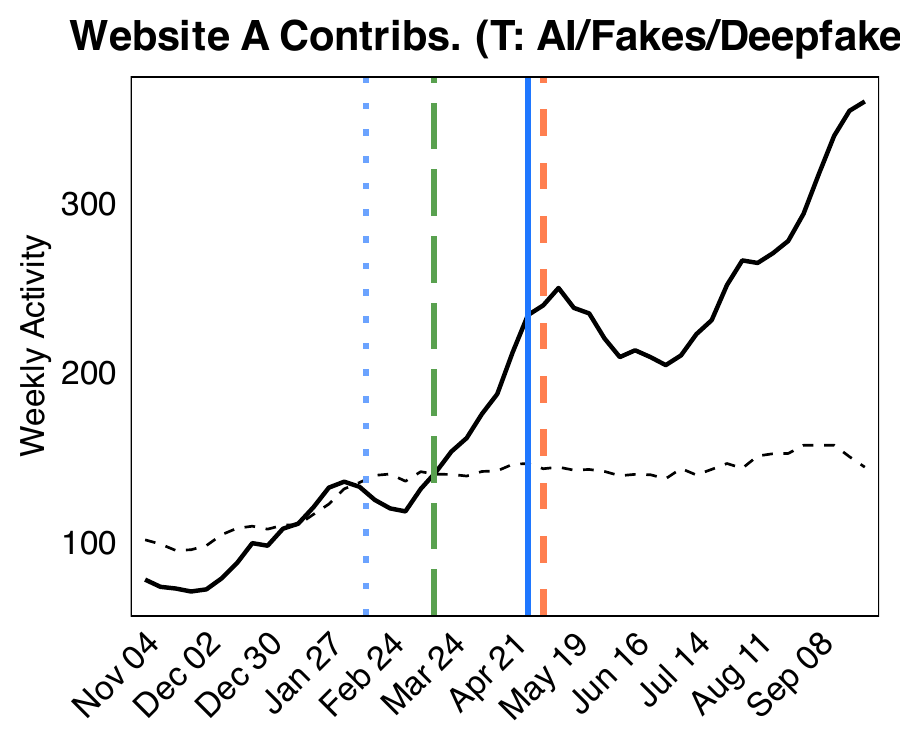}
\includegraphics[width=0.32\textwidth]{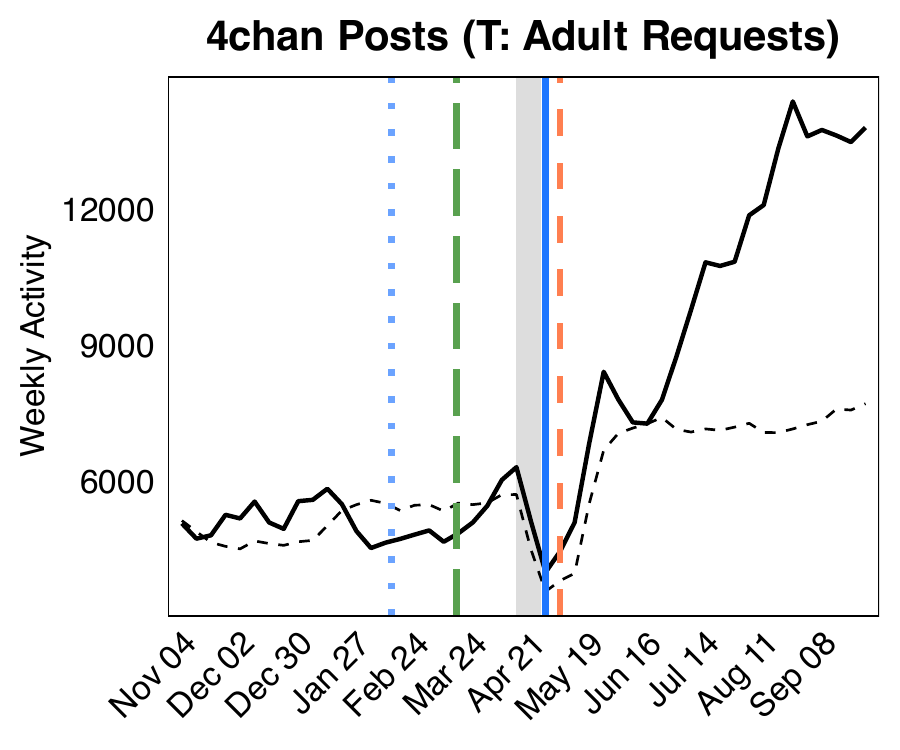}
\\[2mm]
\includegraphics[width=0.85\textwidth]{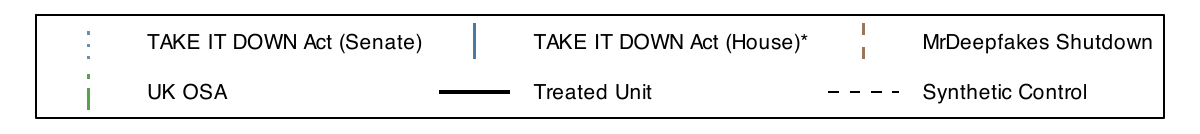}
\caption{\label{fig:main-synth} New posts and newly active contributors for each treated unit across the three websites. Vertical lines represent the passing of the \textit{TAKE IT DOWN} Act at the US Senate, UK OSA, US House of Representatives (main intervention*), and the MrDeepfakes shutdown, in order. Numerical results are shown in Table~\ref{tab:synth-results}. Results for ``Celebrities'' are shown in Figure~\ref{fig:additional}.}
\end{figure*}

\subsection{Synthetic Control Setup}
\label{sec:synth-setup}

To estimate the causal effect of the \textit{TAKE IT DOWN} Act and the subsequent MrDeepfakes takedown on the production of \sncei, we employ the synthetic control method~\cite{abadieUsingSyntheticControls2021}. This method constructs a counterfactual time series (the synthetic control) by creating a linear combination of untreated units---the donor pool or control units---that best reproduces its pre-intervention trajectory. If the counterfactual matches the pre-intervention activity of the treated unit, we assume that in the absence of the intervention, the counterfactual and the treated unit would continue behaving similarly. On the other hand, a divergence between the treated time series (e.g., the activity of a subforum dedicated to \sncei) and the synthetic control (the linear combination of subforums without \ncii) after the intervention represents the estimated treatment effect. All analyses are done using the \texttt{rsynth} R package~\cite{hainmueller2011synth}.

\xhdr{A Primer on Synthetic Controls.}
Synthetic control methods are widely used for policy evaluation in settings with a single treated unit and multiple potential controls. The core premise of the synthetic control method is that an aggregate set of comparison units can reproduce the treated unit's features more accurately than any single donor unit~\cite{abadieUsingSyntheticControls2021}. Additionally, the synthetic control method provides an empirically robust method to select appropriate comparison units. To do this, we identify a convex combination of control units whose pre-treatment outcomes approximate those of the treated unit, thereby generating a baseline for comparison. The post-treatment difference between the treated outcome and its synthetic control provides an estimate of the causal impact of the intervention. The main conditions to be satisfied are that the intervention affects only the treated unit and not the units in the donor pool; that the synthetic control approximates the pre-treatment path; that the synthetic control trend continues post-treatment; and that there are no major structural breaks in the donor pool.

\begin{figure*}[t]
\centering
\includegraphics[width=0.32\textwidth]{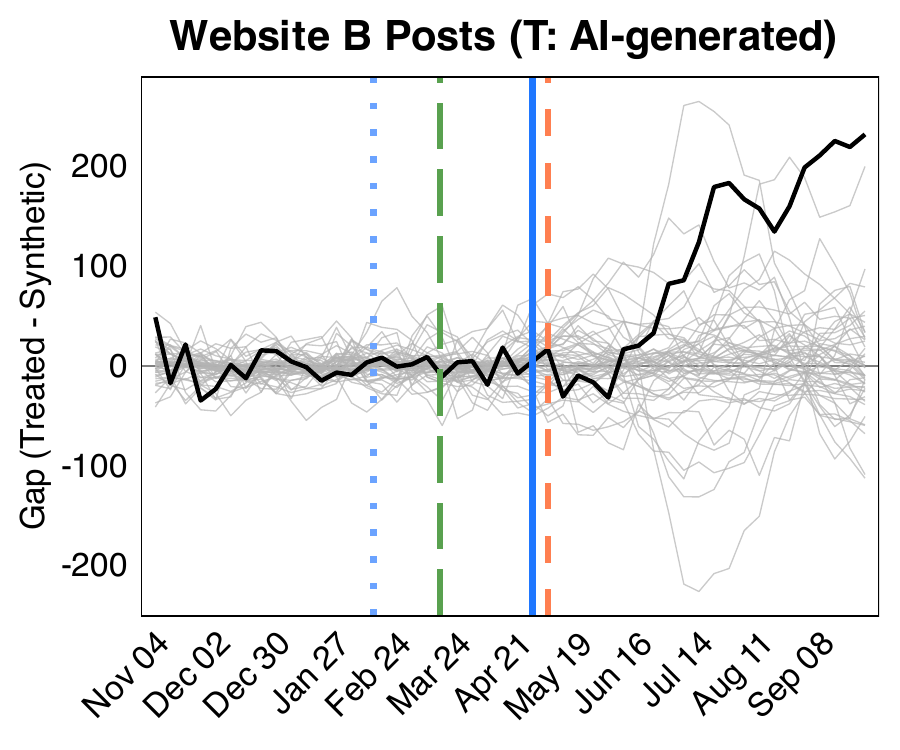}%
\includegraphics[width=0.32\textwidth]{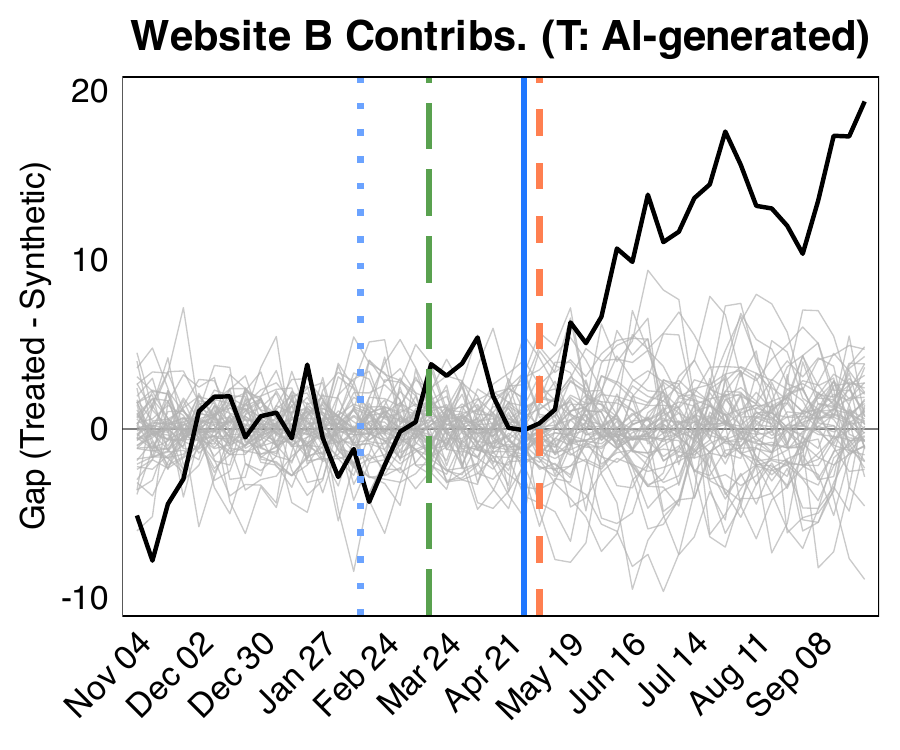}%
\includegraphics[width=0.32\textwidth]{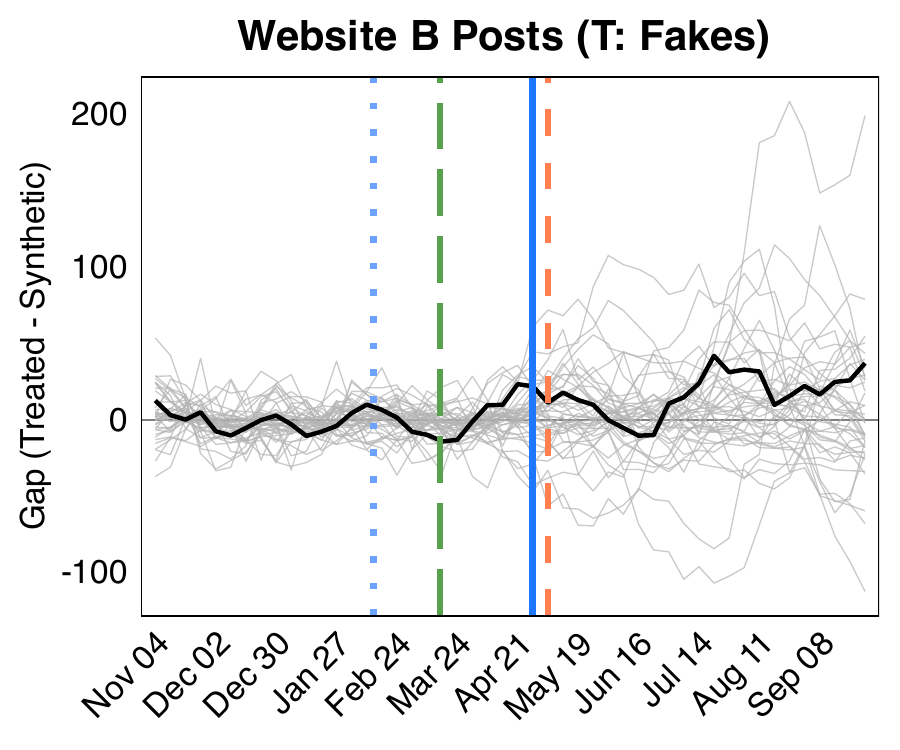}\\[2mm]
\includegraphics[width=0.32\textwidth]{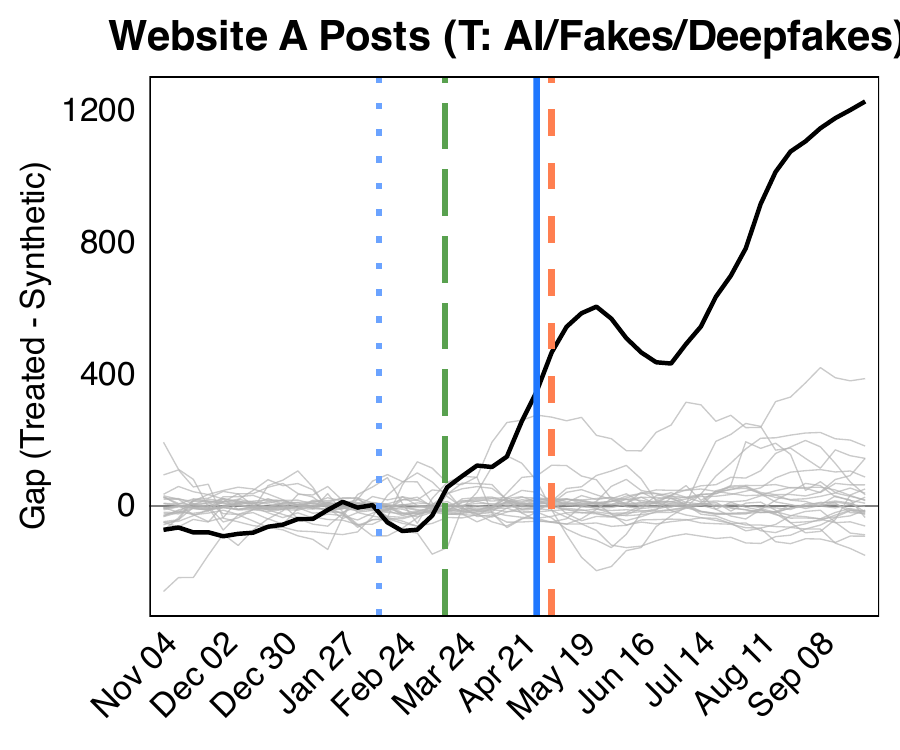}
\includegraphics[width=0.32\textwidth]{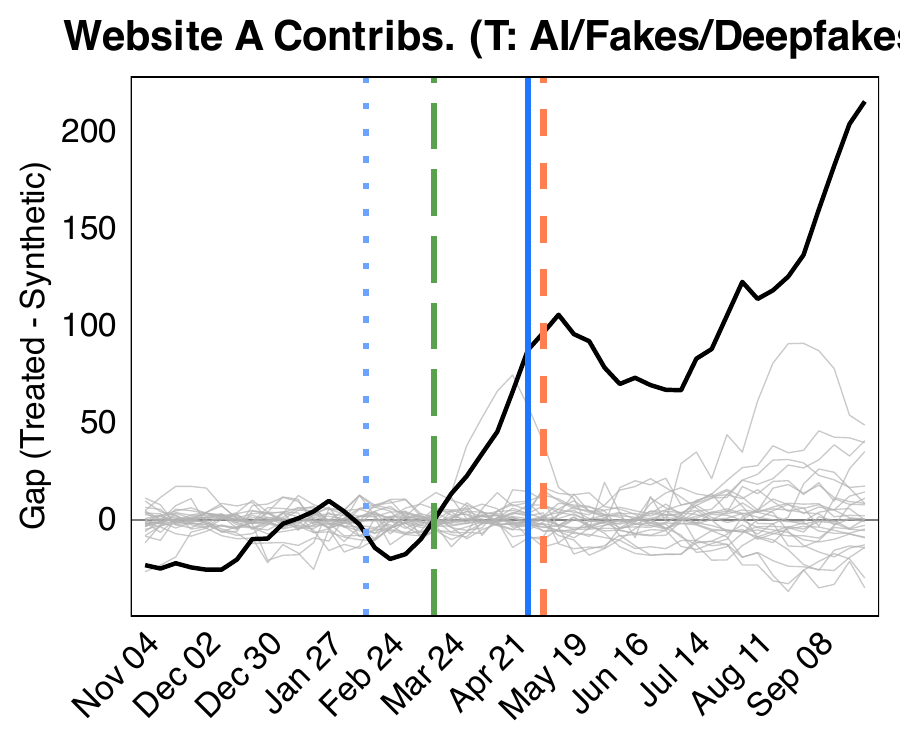}
\includegraphics[width=0.32\textwidth]{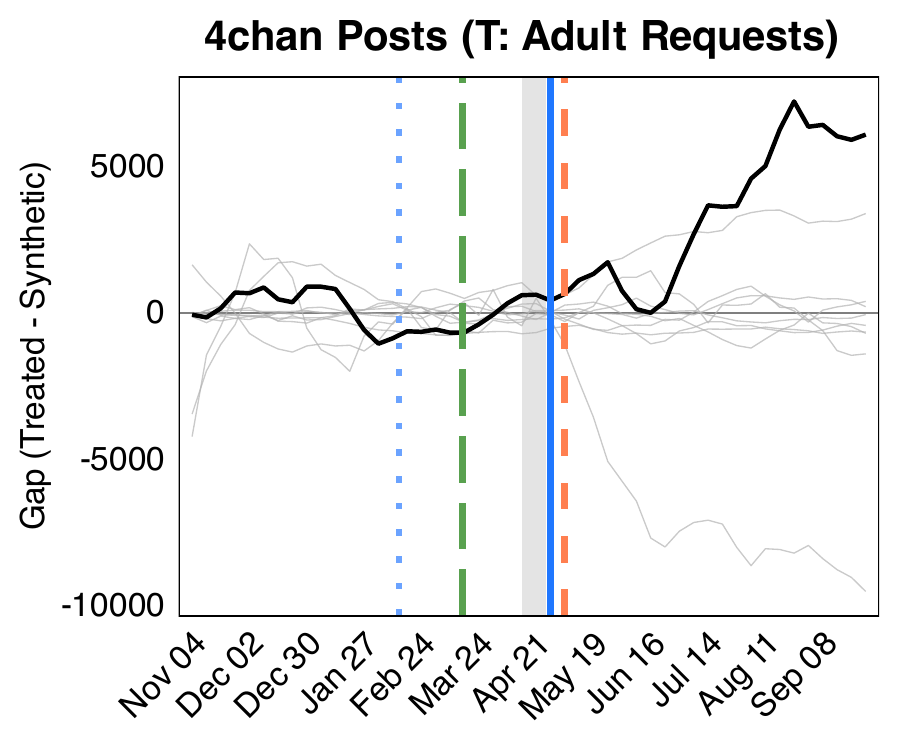}
\\[2mm]
\includegraphics[width=0.85\textwidth]{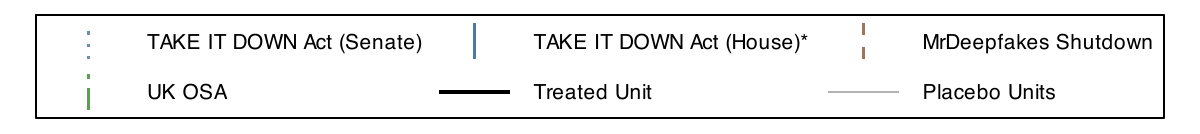}
\caption{\label{fig:placebo-synth} Placebo test results. Solid lines show the difference between each treated unit and its synthetic control. The black line denotes the treated unit; gray lines are possible outcomes for control units. Results for “Celebrities” appear in Figure~\ref{fig:additional}.}
\end{figure*}

\xhdr{Treatment and Units.} We treat the passing of bill S.146, the \textit{TAKE IT DOWN} Act, in the US House of Representatives (April 28th, 2025~\cite{takeitdown_act_2025}),%
\footnote{In light of the President's indicated support~\cite{pressreleaseTakeItAct2025}, we treat April 28 as the point at which the Act became effectively imminent and thus the date of intervention.} 
and the subsequent MrDeepfake shutdown (May 5th, 2025~\cite{wiseMajorDeepfakePorn2025}) as a compound intervention. The treated units are the weekly number of new posts and newly active unique contributors in subforums dedicated to sharing \sncei. We consider a contributor to be newly active if they had no posts in the past 2 months, similar to active editors on Wikipedia~\cite{wikimedia_standard_metrics}. There are 23 weeks or ``post-periods'' after the intervention. For new weekly posts, we build a time series for each subforum on \websiteA, \websiteB, and \fourchan. Similarly, we constructed time series for the number of new weekly contributors to \websiteA and \websiteB. For \websiteA and \websiteB, all posts contribute at least one link, image, or video---but often contain many. For \fourchan, since we treat it as a proxy for requests, we count all posts.

For each time series, we construct donor pools from sexually explicit subforums not oriented toward \sncei (Table~\ref{tab:subforums}). For \websiteA, the treated unit is the ``Fakes/AI/Deepfakes'' subforum; we estimate separate synthetic controls for posts and new contributors using the remaining 32 subforums as donors. For \websiteB, treated units include ``AI-generated'' (posts and contributors), ``Fakes'' (posts), and ``Celebrities'' (posts), with the remaining 65 subforums as donors; we discard contributor-level results for ``Fakes'' and ``Celebrities'' due to poor pre-intervention fit. For \fourchan, we use posts from the ``Adult Requests'' (\texttt{/r/}) subforum as the treated unit and the remaining 13 subforums as donors. We assume the \textit{TAKE IT DOWN} Act primarily affects these treated units, as they disproportionately contain \sncei-related content and contributors.

\xhdr{Statistical Significance}
We assess treatment effects using placebo-based inference based on post- to pre-treatment mean squared prediction error (MSPE) ratios, and visualize effects as deviations between observed outcomes and their synthetic control after the intervention. Because synthetic control methods do not admit conventional sampling-based standard errors or classical p-values, statistical significance is assessed using placebo (permutation) tests~\cite{abadieUsingSyntheticControls2021}. Specifically, we iteratively reassign the treatment to each control unit, re-estimate the synthetic control, and compute the corresponding post/pre-treatment MSPE ratios. The treated unit's ratio rank is then evaluated relative to the empirical distribution induced by these placebo assignments. Note that the minimum attainable permutation p-value in the placebo test is mechanically bounded by the number of available control units. With $N$ total units ($J$ donor units and one treated unit), the smallest possible p-value is $1/(J+1) = 1/N$, corresponding to the treated unit having the most extreme post/pre-treatment MSPE ratio among all placebo assignments. For example, in the case of \fourchan, the donor pool size after filtering implies a minimum achievable p-value of 0.100 (or $1/10)$.

For placebo-based inference, we restrict the donor pool using standard pre-treatment fit and data-quality filters. We retain only units with sufficiently complete and variable pre-treatment activity (at least 80\% of periods observed, non-trivial variation, and repeated nonzero observations), and exclude donors with near-zero activity. We additionally discard placebo units with poor pre-treatment fit, dropping any unit whose pre-treatment MSPE exceeds five times that of the treated unit. This final step can change the number of donors ($N$) across permutations. These restrictions improve comparability and prevent permutation-based inference from being driven by poorly matched controls~\cite{abadie2010california}.

\begin{table*}[htbp]
\centering
\resizebox{\linewidth}{!}{
\begin{threeparttable}
\begin{tabular}{llccccccccc}
\toprule
Website & Metric & Treated Unit & $T_0$\tnote{a} & Direction & Avg. Gap & MSPE Ratio & Rank & N & p-value\tnote{b} & Min. p \\
\midrule
\fourchan & Posts & Requests & 04/28/25 & $+$ & 3,294.2 & 43.4 & 1 & 10 & 0.100\textsuperscript{*} & 0.100 \\
A & Posts & AI/Fakes/Deepfakes & 04/28/25 & $+$ & 737.1 & 79.1 & 1 & 27 & 0.037\textsuperscript{*} & 0.037 \\
A & Contribs. & AI/Fakes/Deepfakes & 04/28/25 & $+$ & 111.2 & 25.8 & 1 & 27 & 0.037\textsuperscript{*} & 0.037 \\
B & Posts & Fakes & 04/28/25 & $+$ & 17.0 & 6.0 & 23 & 43 & 0.535\phantom{*} & 0.023 \\
B & Posts & Celebrities & 04/28/25 & $-$ & -64.8 & 27.1 & 8 & 48 & 0.167\phantom{*} & 0.021 \\
B & Posts & AI-generated & 04/28/25 & $+$ & 102.7 & 68.0 & 1 & 51 & 0.020\textsuperscript{*} & 0.020 \\
B & Contribs. & AI-generated & 04/28/25 & $+$ & 11.1 & 15.4 & 1 & 56 & 0.018\textsuperscript{*} & 0.018 \\
\midrule
\textit{Robustness Checks} &  &   &  &  &  &  &  & &  &  \\
\midrule
A & Posts & AI/Fakes/Deepfakes & 03/16/25 & $+$ & 762.7 & 634.9 & 1 & 27 & 0.037\textsuperscript{*} & 0.038 \\
A & Contribs. & AI/Fakes/Deepfakes & 03/16/25 & $+$ & 102.7 & 201.7 & 1 & 27 & 0.037\textsuperscript{*} & 0.037 \\
\bottomrule
\end{tabular}
\end{threeparttable}
}
\begin{tablenotes}
\item[*] * indicates p $<$ 0.05 or p $=$ min. p.
\end{tablenotes}
\caption{\label{tab:synth-results}
Synthetic control and placebo results. Backdating robustness check is reported for \websiteA. $T_0$ represents intervention date. Direction: $+$ indicates an increase and $-$ a decrease in the outcome. Avg. Gap denotes the mean post-treatment difference (Treated $-$ Synthetic) per time unit. Reported p-values are permutation p-values based on the treated unit's rank in the MSPE placebo distribution. Min. p is the minimum possible p-value implied by the placebo sample size ($1/N$). 
}
\end{table*}

\section{Results}
\label{sec:results}

We present the graphical comparison of each treated unit against its counterfactual in Figure~\ref{fig:main-synth}. The graphical results of the placebo tests, in which we perform a synthetic control analysis for each unit as a treated unit and compute the gap between the treated and the counterfactual, are shown in Figure~\ref{fig:placebo-synth}. The numerical results are shown in Table~\ref{tab:synth-results}. Lastly, the list of subforums and their activity is shown in Table~\ref{tab:subforums}.

\subsection{\websiteA}
For \websiteA, we find that the number of posts and newly active contributors in the ``AI/Fakes/Deepfakes/'' subforum significantly increased after the passing of the bill and the shutdown of the MrDeepfakes website (Figure~\ref{fig:main-synth}). Both the number of new contributors and the number of new posts increased substantially following the compound intervention. The post-treatment MSPE exceeds the pre-treatment MSPE, indicating decoupling from the counterfactual. Further, the ratio of post- over pre-treatment MSPE is larger than when any other subforum is tested as a treated unit (i.e., rank 1 in the permutation test). Visually, the placebo tests also confirm a clear increase in the number of posts and contributors as compared to other subforums (Figure~\ref{fig:placebo-synth}).

The average gap over the synthetic control was of 737.1 more posts and 111.2 more contributors per week after the intervention. Notably, this subforum was lagging in activity as compared to the synthetic control between November and December 2024. We begin observing a break from the trend mid-March and by mid-April the gap is apparent. Visually, the increase in posts and users seems to begin before the intervention, which may indicate an anticipation effect or the impact of another event.

\subsubsection{Backdating Robustness Checks for Anticipation.}
In both cases (posts and contributors), outcomes diverge \textit{prior} to the bill's passage in the House, motivating a robustness check for potential anticipation effects. Backdating the intervention in a synthetic control framework is a common robustness check when anticipation effects seem to exist. This procedure does not mechanically bias estimated treatment effects, as the method places no restriction on the timing or constancy of post-intervention impacts~\cite{abadieUsingSyntheticControls2021}. We re-estimate the synthetic control using the UK's OSA deadline platform compliance (March 16th, 2025). There are 29 weeks after the intervention. We find that the average post-intervention gap marginally increases from 737.1 to 762.7 for new posts. On the other hand, the average gap for the number of new contributors per week remains stable (111.2 to 102.7). However, in both cases the corresponding MSPE ratios increase substantially: from 79.1 to 634.9 for posts and from 25.8 to 201.7 for contributors (shown in Table~\ref{tab:synth-results}).

The robustness check cannot tell us whether the earlier divergence is driven by anticipation of the \textit{TAKE IT DOWN} Act or by the UK OSA's enforcement deadline. Because we cannot separately identify exposure to each law, shifting the treatment date earlier does not isolate a distinct UK OSA effect. Instead, it provides a plausible alternative explanation. Importantly, the estimated post-\textit{TAKE IT DOWN} Act divergence persists under alternative timing assumptions, indicating that the main effect is robust, even if its precise onset cannot be attributed to a single policy date.

\begin{findingbox}
\textbf{Result.} We find a significant increase in posting activity following the \textit{TAKE IT DOWN} Act. At the same time, outcomes begin to diverge before the bill’s passage, consistent with anticipation effects or heightened regulatory salience during this period (including contemporaneous enforcement of UK's OSA). Our results, therefore, suggest that the Act's effects on \websiteA occurred within an ongoing policy window rather than from a single shock.
\end{findingbox}

\subsection{\websiteB}
We observe a significant increase in posts and users in the ``AI-generated'' category after the intervention (Figure~\ref{fig:main-synth}). For posts, there is a small decrease until mid-June but then a sharp and continuous increase until mid-July. For contributors, the increase seems to immediately increase after the interventions. For posts tagged as ``AI-generated'', the MSPE ratio is the highest in the permutation distribution, with p-value 0.020 ($1/51$). We observed an average gap of 102.7 more posts per week as compared to its synthetic control. For contributors of ``AI-generated'' posts, the MSPE ratio rank is also the highest, with p-value 0.018 ($1/56$), and the average gap was of 11.1 more new contributors per week than the synthetic control for the study period. Visually, these results can be seen in Figure~\ref{fig:placebo-synth}. In the ``Fakes'' subforum, the weekly number of posts stayed flat throughout the study period relative to the synthetic control. This result indicates that activity continued to trend upwards and was not significantly affected by the intervention . On the other hand, in the ``Celebrities'' subforum, we observe a seeming decrease in posts as compared to its synthetic control. While the MSPE ratio does not indicate a significant change post-intervention, celebrity-related posts trended downwards post-intervention. We note that, contrary to \websiteA, \websiteB has less robust content guidelines which can affect the quality of the proxy; we discuss this limtation in Section~\ref{sec:limitations}. Furthermore, posts can overlap across categories, which means our average gap estimates across categories are not additive. As mentioned in Section~\ref{sec:synth-setup}, we did not achieve a good pre-intervention fit for the number of weekly contributors for the ``Fakes'' and ``Celebrities'' subforums with their corresponding synthetic controls, thus we omitted these from analysis.

\begin{findingbox}
\textbf{Result.} The passage of the \textit{TAKE IT DOWN} Act caused a significant increase in the number of posts and contributors in the ``AI-generated'' subforum on \websiteB. We found no significant change post-intervention in the number of explicit posts tagged as ``Fakes'' and those tagged with ``Celebrities.'' The number of ``Fakes'' posts continued to have a positive slope after bill's passage, similar to other subforums in \websiteB, while the ``Celebrities'' posts had a negative slope after the bill's passage.
\end{findingbox}

\subsection{\fourchan}
For \fourchan, we find a significant increase in posts after the intervention, as observed in Figure~\ref{fig:main-synth}. Importantly, \fourchan faced an incident that took the platform down between April 14th and April 25th, 2025. Three days later, the \textit{TAKE IT DOWN} Act passed the US House of Representatives. As shown, both the number of posts in the ``Adult Requests'' subforum and the synthetic control steadily return to the pre-intervention trend around May 19th. However, while the number of posts in the ``Adult Requests'' subforum lagged the synthetic control prior to the incident on April 14th, it rapidly surpassed the synthetic control after April 28th and continues to rise through May. On average, the ``Adult Requests'' subforum received 3,294.2 more posts than its synthetic control. The MSPE ratio for the ``Requests'' subforum was also the highest (see Figure~\ref{fig:placebo-synth} and Table~\ref{tab:synth-results}), resulting in a p-value of $0.100$ ($1/10$). This means that the passing of the \textit{TAKE IT DOWN} Act by the US House of Representatives and the MrDeepfakes shutdown caused an increase in the number of posts in the \fourchan ``Adult Requests'' subforum. As we show below, these posts are primarily requests for \sncei.

\subsubsection{Requests for \sncei}
As discussed in Section~\ref{sec:background}, posts on the Adult Requests board (\texttt{/r/}) have historically contained requests for \sncei, using tools such as Adobe Photoshop in the past and, more recently, AI. For example, users post requests asking others to use AI for: nudification, face swapping (e.g., overlapping someone's face on someone performing a sexual act), body part alteration (e.g., enlarging breasts), or to create sexually explicit clips using pictures and prompts they provide. Other users will then fulfill those requests and post the requested content. 

To quantify the prevalence of these requests, we randomly sampled metadata (post title and body text) from 1,000 original posts (i.e., the first post in a thread) posted during our study period. We manually labeled whether each post was a request for \sncei; however, because we did not collect or use images for this labeling procedure, we may have missed contextual cues. We found that 86.1\% of the posts we sampled were requests for \sncei. This does not include posts that may constitute \ncii but that do not use AI, such as requests of ``leaks'' (e.g., content that was obtained without authorization). We note that some threads have dozens of media replies, which constitute either the pictures of targets or the results of the requests. Victims were often described as girlfriends, wives, coworkers, and classmates. The majority of cases seemed to concern private individuals, rather than public figures (e.g., influencers, celebrities, or other people of public renown). While a comprehensive characterization of this subforum is beyond the scope of this paper, these results help qualify the results of our synthetic control tests.

\begin{findingbox}
\textbf{Result.} The passage of the \textit{TAKE IT DOWN} Act increased the number of posts in the ``Adult Requests'' subforum. During our study period, 86.1\% of posts in this subforum were requests for \sncei. 
\end{findingbox}

\section{Discussion}
\label{sec:discussion}

The closely timed regulatory and platform shocks we study plausibly could have deterred \sncei production and sharing. 
The \textit{TAKE IT DOWN} Act represented a major federal milestone, expanding criminal liability for \sncei and introducing a notice-and-takedown regime that could increase both perceived risk and removal capacity. 
In parallel, the UK’s Online Safety Act (OSA) signaled heightened enforcement of platform duties around illegal harms. 
Consistent with the possibility of deterrence, MrDeepfakes shut down roughly one week after the Act’s passage (though the causal link is not confirmed), and major AI generation platforms such as CivitAI---long associated with \sncei creation~\cite{maibergCivitaiOctoMLIntroduce2023}---updated their policies in anticipation of increased scrutiny~\cite{civitaiPolicyContentAdjustments}.

Yet our results suggest that, rather than suppressing participation, regulatory pressure primarily displaced activity across platforms, with no evidence of a slowdown in overall activity. This migration was neither uniform nor solely driven by the shutdown of MrDeepfakes. \websiteA experienced a surge in activity as early as March, consistent with anticipation effects in which users appear to track policy developments and adjust their platform choices in advance, whereas \websiteB and \fourchan saw increases only after the intervention window. These patterns indicate that regulatory pressure shaped platform choice and the timing of participation, but not underlying demand. Beyond displacement, we observe clear net growth: over our one-year study period, all three websites recorded more posting and contributor activity than MrDeepfakes did over its seven-year lifetime. Moreover, in five of seven treated units, \sncei subforums grew faster than other subforums on the same platforms. Because our analysis measures contributor activity rather than consumption, these increases likely represent a lower bound on total engagement: rising requests for \sncei suggest that audience size and demand expanded well beyond what contributor counts alone capture.


These results echo outcomes of the anti-sex trafficking bills FOSTA and SESTA. In 2018, two weeks after the US Congress passed these bills, two large online sex advertising platforms ceased operation~\cite{zengInternetGovernanceSite2022}. However, as \citet{zengInternetGovernanceSite2022} find, the bills' passage and the riddance of these two major websites had seemingly no impact on prostitution arrests and violence against women~\cite{zengInternetGovernanceSite2022}. Instead, within months, there was a significant increase in the number of commercial sex advertisements and visitors to offshore sites not subject to US legal jurisdiction~\cite{zengInternetGovernanceSite2022}. Scholars have found similar results when studying the deplatforming of extremist sites~\cite{vu2024no}, online darkweb marketplaces~\cite{soskachristin2015}, and cybercriminal services~\cite{vu2025assessing}. Lest we forget, MrDeepfakes itself launched just a week after Reddit banned the \texttt{/r/deepfakes} subreddit~\cite{timmermanStudyingOnlineDeepfake2023}. 

\subsection{Limitations}
\label{sec:limitations}

\xhdr{Donor Unit Comparability.} Synthetic control designs depend on the availability of comparable donor units and stable pre-treatment dynamics. We argue that donor units are comparable insofar as they coexist on the same platforms as the treated units, but contain content that should be differentially affected by the intervention. Nevertheless, as with all observational designs, residual confounding and sensitivity to design choices cannot be fully ruled out.

\xhdr{Construct Validity.} Our goal is to study the dissemination of \sncei. To do so, we employ proxies based on the number of posts and contributors across three websites. \websiteA has a clearly defined subforum dedicated to \sncei with explicit content guidelines and activity thresholds that constrain where users may post or initiate threads. These features make \websiteA a reliable proxy for measuring activity related to the sharing of \sncei. By contrast, \websiteB relies on user-generated categorization across three categories of interest, and newly registered users may upload content immediately. While \websiteB enforces general platform rules (e.g., prohibitions on underage content), it does not provide category-specific guidance. As a result, although \websiteB remains a useful site for studying \sncei dissemination, we cannot rule out the possibility that observed changes reflect shifts in users’ tagging behavior (e.g., avoiding the ``Celebrities'' or ``Fakes'' labels) rather than changes in the underlying volume of \sncei content. In the case of \fourchan, we estimate the proportion of \sncei-related requests using textual analysis and manual labeling.

\xhdr{Causal Effect of the Interventions.} Synthetic control analyses typically focus on a single intervention. In our case, the Act's passage plausibly raises expected enforcement and compliance costs, deterring \sncei creation and circulation. The period, however, also includes other relevant developments, notably the UK's OSA enforcement deadline and the shutdown of MrDeepfakes. For \websiteA, outcomes diverge prior to the Act's passage, consistent with anticipation effects or UK OSA exposure (which we assess in a robustness check). By contrast, other platforms exhibit no comparable pre-trends. Together, these events define a broader policy window shaping \sncei activity. Although platforms were exposed to similar regulatory signals, their responses differ, likely reflecting variation in perceived enforcement risk or relevance. While our design cannot isolate individual effects, the magnitude and persistence of post-intervention deviations support a causal interpretation tied to heightened regulatory pressure and deplatforming.

\subsection{Future Work}
\label{sec:future-work}
Our findings suggest that future work should move beyond treating deplatforming or monetization disruption as self-contained deterrents and instead examine how users adapt to regulatory and infrastructural pressure. The uneven timing of activity increases across websites suggests that migration may be shaped by platform affordances and expectations about enforcement risk. Notably, many of the platforms we study provide free access to \sncei content, and a meaningful share of participation (such as bespoke content requests on \fourchan) appears to be sustained without explicit financial incentives. 

Future research should therefore examine not only economic motivations, but also social, reputational, and ideological drivers that sustain participation. While prior work emphasizes targeting payment processors, advertising networks, search engines, or infrastructure providers, adaptive responses---such as shifts to cryptocurrency payments or reliance on alternative service providers~\cite{civitaiCreditCardPayments,Han_Kumar_Durumeric_2022,papadogiannakisWelcomeDarkSide2025}---suggest that these strategies may be easily circumvented. Moreover, the continued growth of dark-web marketplaces despite substantial access barriers (e.g., limited discoverability and technical expertise requirements) indicates that search engine deprioritization or pushing content to harder-to-access venues may have limited deterrent effects. Finally, future work should investigate why recent legal interventions appear to have weak deterrent effects, including whether users expect enforcement to be avoidable, or whether the absence of sexual education and moral condemnation of non-consensual image abuse leaves these laws without the urgency required to change behavior.

{\small
\bibliography{aaai25}
}

\appendix
\clearpage

\section{Appendix}
\subsection{Terminology}
\label{apx:terminology}

We use non-consensual intimate imagery (\ncii) as the umbrella term to refer to all forms of non-consensual intimate content (whether synthetic or not). \ncii is a form of image-based sexual abuse (\ibsa), a term that covers the non-consensual creation or distribution of private sexual images~\cite{mcglynn2016not}. This term was adopted as a replacement for ``revenge porn,'' frequently criticized by scholars~\cite{mcglynn2016not,franksCriminalization2024}. 
Within \ncii, we adopt the term synthetic non-consensual explicit imagery (\sncei) to refer more precisely to synthetic content that is sexually explicit. We pay special attention to \sncei created with AI tools (e.g., nudifiers; \citet{gibsonAnalyzingAINudification2025}).

We note the absence of definitional consensus across the literature. Different terms are used to change the scope (e.g., intimate vs. explicit), intention (e.g., using ``pornography'' to emphasize the prurient nature~\cite{franksCriminalization2024}), or to add emphasis to the production method (e.g., ``AI-generated''). Past work has various terms including AI-IBSA~\cite{umbachNonConsensualSyntheticIntimate2024}, AI-NCII (AI-generated non-consensual intimate imagery;~\citet{trifonovaMisinformationFraudStereotyping2024}), 
NSII (non-consensual synthetic intimate imagery;~\citet{umbachNonConsensualSyntheticIntimate2024}), SNEACI (synthetic non-consensual explicit AI-created imagery;~\citet{gibsonAnalyzingAINudification2025}), NCID (non-consensual intimate deepfakes;~\citet{kiraWhenNonconsensualIntimate2024}),
deepfake NCII~\cite{hanCharacterizingMrDeepFakesSexual2025}, non-consensual deepfakes~\cite{timmermanStudyingOnlineDeepfake2023}, and colloquially deepfake nudes/pornography~\cite{szetoThisCanadianPharmacist2025,wiseMajorDeepfakePorn2025,teamUnmaskingMrDeepFakesCanadian2025}.
All these terms could be applied to the content distributed on the websites we study---that we refer to as \sncei.

\subsection{List of Subforums per Website}
\label{apx:subforums}

We provide the full list of subforums for each website and highlight the treated units and donor units in Table~\ref{tab:subforums}. For each subforum, we display the number of posts and contributors to contextualize the activity of each subforum during our study period.

\subsection{Perceived Legal Compliance per Website}

\xhdr{\websiteA.} To the best of our knowledge, \websiteA does not honor Digital Millennium Copyright Act (DMCA) takedown requests, nor \textit{TAKE IT DOWN Act} requests, nor it provides any formal mechanism for content removal. The platform displays a brief and generic privacy policy and terms of service. While the applicable legal jurisdiction remains unclear, domain registration and hosting records across multiple mirrors consistently indicate infrastructure associated with Belize, alongside the use of multiple domain registrars, rotating DNS providers, and various companies as registrants.

\xhdr{\websiteB.} \websiteB seems to generally comply with takedown requests, at least retroactively. The platform maintains a growing list of sites, models, and publishers that have requested their material not be hosted and claims to actively monitor and remove such content. At the time of writing, this list includes over 700 entries. The website also provides a DMCA notice with contact information, as well as a separate GDPR-compliant privacy policy that identifies a responsible contact person. The site displays a ``2257'' compliance statement pursuant to 18~U.S.C.\S2257\cite{usc18sec2257}. It also links to external resources addressing CSAM, including ASACP and Offlimits NL, a Dutch organization. Additionally, domain registration records indicate a Dutch registrar. The site's administrators have stated that they will comply with the UK's OSA age verification guidelines, but encourage users to use a VPN. While these indicators are not conclusive, taken together they suggest that \websiteB operates with awareness of EU, UK, and US compliance regimes and presents itself as a more compliant platform than \websiteA.

\xhdr{\fourchan.} \fourchan has historically operated in a gray area due to its lax content moderation and anonymity. The website has a history of questionable content on a variety of fronts, ranging from extremist and hateful discourse to \ncii. \fourchan displays US-based contact information and business registration. They also provide information on how to submit DMCA and \textit{TAKE IT DOWN} requests. However, compliance for content removal requests on \fourchan is tricky. Since posts are ephemeral and disappear within days of submission, takedown requests are likely too slow to be effective. By the time they are submitted and serviced, the post in question may have already been deleted. \fourchan was among the first websites to be contacted (and then fined) by Ofcom as part of their enforcement program~\footnote{\fourchan has challenged the enforcement. The legal dispute is currently developing.~\cite{4chanChallengesOfcoms}}~\cite{ofcom_4chan_investigation_2025}. These events indicate that \fourchan seems to comply with US laws following a notice-and-takedown approach for most content, except CSAM. It also operates with awareness (but not necessarily in compliance) of UK legislation.

\begin{figure*}[ht!]
    \centering
    \includegraphics[width=0.32\linewidth]{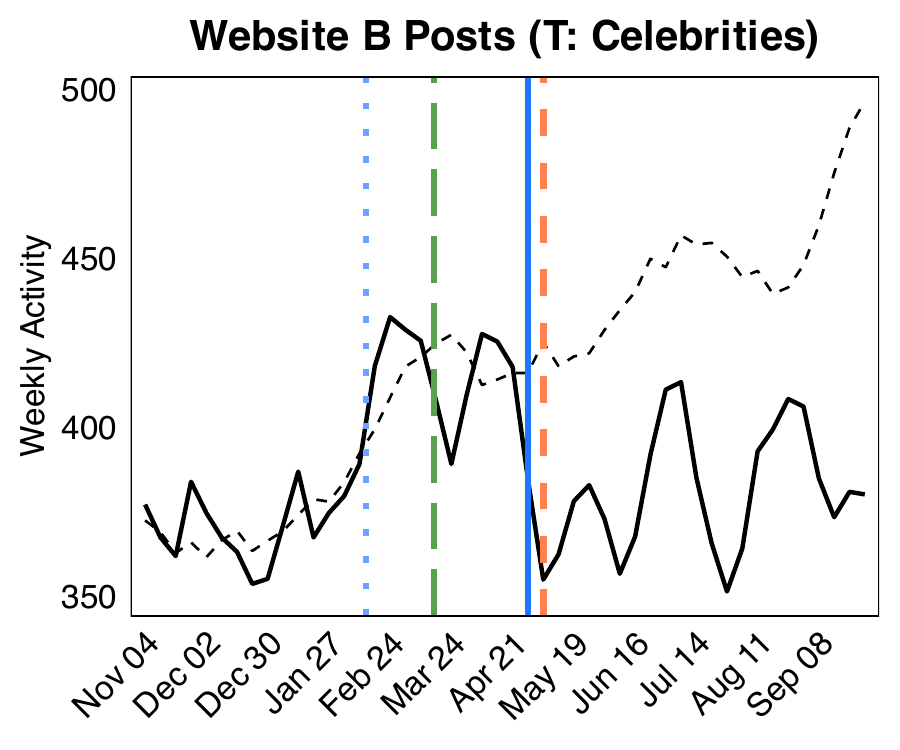}
    \includegraphics[width=0.32\linewidth]{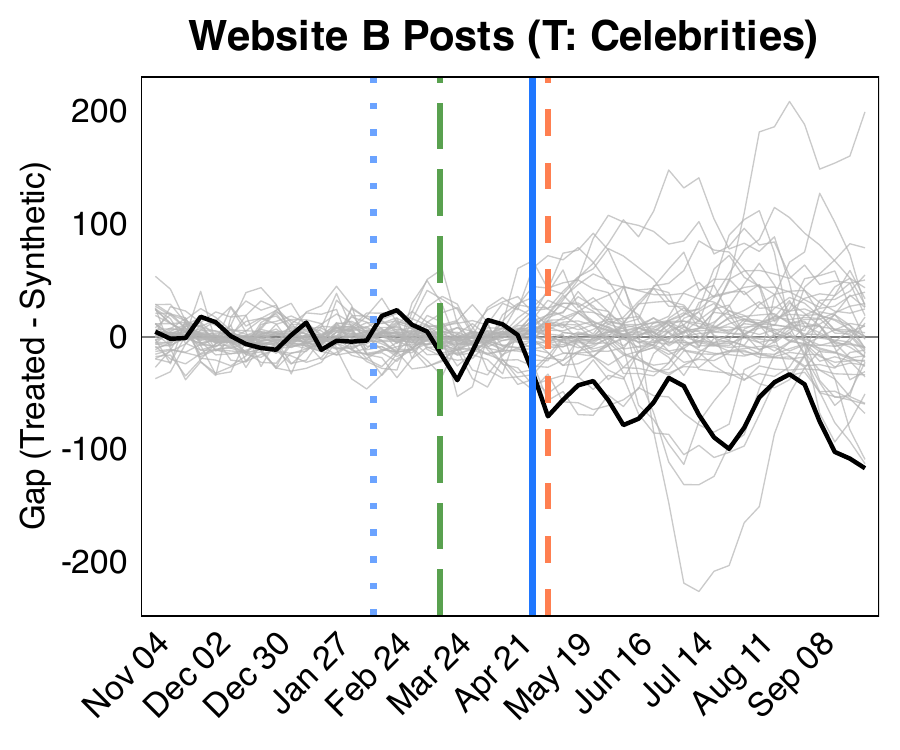}
    \caption{Continued from Figures~\ref{fig:main-synth} and ~\ref{fig:placebo-synth}, graphs corresponding to ``Celebrities'' posts on \websiteB. }
    \label{fig:additional}
\end{figure*}

\subsection{Additional Figures}

For presentation purposes, the figures corresponding to the synthetic control and placebo tests for ``Celebrities'' posts from \websiteB are shown here, see Figure~\ref{fig:additional}).

\begin{table*}[ht]
\renewcommand{\arraystretch}{1.25}
\centering
\begin{threeparttable}
\begin{tabular}{l p{15.45cm}}
\hline
Website & Subforums (Nr. of Posts / Nr. of Contributors). \\
\hline
\parbox[t]{1.5cm}{Website A\\$n=32$} &
ASMR (9,630 \textbar{} 2,034), 
Asians (51,331 \textbar{} 8,211), 
Cam Girls (36,353 \textbar{} 6,023), 
Celebrities\textsuperscript{a} (86,738 \textbar{} 6,071), 
\textbf{Fakes/AI/Deepfakes} (75,297 \textbar{}10,941), 
Hotwives + QoS (11,018 \textbar{} 2,961), 
Instagram (141,203 \textbar{} 20,250), 
ManyVids (13,578 \textbar{} 3,375), 
Onlyfans (235,976 \textbar{} 31,552), 
Patreon (19,698 \textbar{} 3,596), 
Professional Modelling Sites (17,824 \textbar{} 1,598), 
Reddit (14,395 \textbar{} 4,185), 
Suicide Girls (563 \textbar{} 167), 
TikTok (30,478 \textbar{} 6,962), 
Twitch (46,658 \textbar{} 6,618), 
XXX/Porn (89,731 \textbar{} 10,316), 
YouTube (12,205 \textbar{} 3,014), 
(Brazilians) General Chat (8,673 \textbar{} 2,391), 
(Brazilians) Close Friends (4,952 \textbar{} 1,079), 
(Brazilians) Photo Essays (4,858 \textbar{} 292),  
(Brazilians) Onlyfans and Patreon (8,549 \textbar{} 2,440), 
(Brazilians) Other Paid Content (7,414 \textbar{} 1,901), 
(Brazilians) Requests (14,866 \textbar{} 4,728), 
(Brazilians) Privacy (7,110 \textbar{} 2,049), 
(Brazilians) Transgender General (4,548 \textbar{} 1,240), 
(Brazilians) Transgender Requests (2,676 \textbar{} 882), 
(Brazilians) Celebrities\textsuperscript{a} (13,791 \textbar{} 957), 
(Brazilians) VIP (1,762 \textbar{} 238), 
(Transgender) General (27,776 \textbar{} 4,809), 
(Transgender) Model Discussion (12,039 \textbar{} 2,479), 
(Transgender) Requests (38,357 \textbar{} 8,512), 
(Other) General (8,080 \textbar{} 2,014), 
\textit{Total: 1,031,139 \textbar{} 157,914.}\\ 

\hline
\parbox[t]{1.5cm}{Website B\\ $n=68$} &
Lesbian (8,428 \textbar{} 1,130), 
Amateur (213,260 \textbar{} 15,766), 
Anal (21,753 \textbar{} 2,936), 
Anime/Cartoon (21,519 \textbar{} 1,956), 
Asian (24,460 \textbar{} 2,366), 
BBW (21,404 \textbar{} 2,519), 
Black/Ebony (11,023 \textbar{} 1,567), 
Big Tits (61,442 \textbar{} 5,582), 
Bondage/S\&M (24,038 \textbar{} 2,360), 
Vintage (13,127 \textbar{} 1,048), 
Cumshot (15,883 \textbar{} 2,773), 
Double Penetration (2,307 \textbar{} 430), 
Fetish (52,183 \textbar{} 4,941), 
Redhead (7,265 \textbar{} 1,256), 
Gang Bang (3,281 \textbar{} 680), 
Gay (14,647 \textbar{} 2,122), 
Interracial (15,876 \textbar{} 1,923), 
Latino/Latina (11,554 \textbar{} 1,378), 
Masturbation (13,296 \textbar{} 2,195), 
Mature (64,251 \textbar{} 5,074), 
Miscellaneous (30,762 \textbar{} 2,401), 
Oral (11,250 \textbar{} 1,717), 
Orgy/Groupsex (3,409 \textbar{} 564), 
CD/TV (22,080 \textbar{} 2,491), 
Asses (37,400 \textbar{} 4,111), 
Computer Generated (6,406 \textbar{} 786), 
Voyeur (19,002 \textbar{} 2,275), 
Teen (48,277 \textbar{} 4,410), 
Hardcore (37,489 \textbar{} 2,311), 
Softcore (41,154 \textbar{} 2,596), 
\textbf{Celebrities} (23,688 \textbar{} 2,033), 
Pregnant (2,974 \textbar{} 706), 
Swingers (3,379 \textbar{} 690), 
Insertion (4,167 \textbar{} 745), 
Feet (17,955 \textbar{} 1,666), 
Lactating (444 \textbar{} 171), 
Gothic (1,589 \textbar{} 470), 
Arabian (1,960 \textbar{} 410), 
Hairy (11,948 \textbar{} 1,566), 
Upskirt (4,980 \textbar{} 808), 
Flashing (13,289 \textbar{} 1,823), 
Outdoors (16,443 \textbar{} 1,946), 
Downblouse (1,636 \textbar{} 397), 
Funny/Oops (2,972 \textbar{} 652), 
Blondes (23,095 \textbar{} 3,110), 
Pornstars (25,890 \textbar{} 1,766), 
Screencaps (2,235 \textbar{} 398), 
\textbf{Fakes} (10,566 \textbar{} 1,362), 
Squirting (919 \textbar{} 249), 
Bizarre (6,509 \textbar{} 1,203), 
Bukkake (1,846 \textbar{} 519), 
Captions (30,013 \textbar{} 2,881), 
Crossdressing (18,040 \textbar{} 2,396), 
Panties (11,669 \textbar{} 1,830), 
Big Cocks (23,469 \textbar{} 3,221), 
CFNM (1,476 \textbar{} 362), 
Gaping/Stretching (4,323 \textbar{} 758), 
Homemade (39,732 \textbar{} 5,000), 
Facial (7,817 \textbar{} 1,452), 
Creampie (3,163 \textbar{} 860), 
Handjob (2,852 \textbar{} 744), 
Filthy (5,534 \textbar{} 820), 
Shemale (13,257 \textbar{} 1,672), 
Animated GIFs (32,777 \textbar{} 2,805), 
Uniforms (3,962 \textbar{} 902), 
Granny (12,418 \textbar{} 1,252), 
Old Men (2,834 \textbar{} 677), 
\textbf{AI-generated} (43,981 \textbar{} 3,715). 
\textit{Total: 1,314,027 \textbar{} 137,701.}
\\
\hline
\parbox[t]{1.5cm}{\fourchan\tnote{b} \\ $n=13$} &
Adult Cartoons (411,812),
Hentai Alt. (174,014),
Ecchi (67,034),
Adult GIF (1,795,200),
Hentai (333,173),
Hardcore (58,945),
Handsome Men (26,461),
High Resolution (138,577),
\textbf{Adult Requests} (496,531),
Sexy Beautiful Women (197,873),
Torrents (21,422),
Yuri (88,838),
Yaoi (11,781).
\textit{Total: 3,821,661.} \\
\hline
\end{tabular}
\begin{tablenotes}
\item[a] This subforum is dedicated to non-fake pictures and videos from celebrities, typically obtained from movies, photoshoots, and magazines.
\item[b] Number of contributors not available because authors are anonymous.
\end{tablenotes}
\end{threeparttable}
\caption{\label{tab:subforums} Included subforums per website with the number of unique posts and contributors for each subforum between October 1st, 2024 and September 30th, 2025 (our study period). Bolded subforums are treated units, unbolded subforums constitute the donor pool. All subforums are dedicated to adult content.}
\end{table*}

\end{document}